\documentclass[prb,reprint,notitlepage,amsmath,amsfonts,amssymb,superscriptaddress]{revtex4-1}
\usepackage{graphicx}
\usepackage{bm}
\usepackage{color}

\begin{document}
\title{Projected Coupled Cluster Theory: Optimization of cluster amplitudes in the presence of symmetry projection}
\author{Yiheng Qiu}
\affiliation{Department of Chemistry, Rice University, Houston, TX 77005-1892}

\author{Thomas M. Henderson}
\affiliation{Department of Chemistry, Rice University, Houston, TX 77005-1892}
\affiliation{Department of Physics and Astronomy, Rice University, Houston, TX 77005-1892}

\author{Jinmo Zhao}
\affiliation{Department of Chemistry, Rice University, Houston, TX 77005-1892}

\author{Gustavo E. Scuseria}
\affiliation{Department of Chemistry, Rice University, Houston, TX 77005-1892}
\affiliation{Department of Physics and Astronomy, Rice University, Houston, TX 77005-1892}
\date{\today}

\begin{abstract}
Methods which aim at universal applicability must be able to describe both weak and strong electronic correlation with equal facility.  Such methods are in short supply.  The combination of symmetry projection for strong correlation and coupled cluster theory for weak correlation offers tantalizing promise to account for both on an equal footing.  In order to do so, however, the coupled cluster portion of the wave function must be optimized in the presence of the symmetry projection.  This paper discusses how this may be accomplished, and shows the importance of doing so for both the Hubbard model Hamiltonian and the molecular Hamiltonian, all with a computational scaling comparable to that of traditional coupled cluster theory.
\end{abstract}

\maketitle

\section{Introduction}

Single-reference coupled cluster theory\cite{Coester1958,Cizek1966,Paldus1999,Bartlett2007,ShavittBartlett} is often considered the gold standard of quantum chemistry. It is conceptually straightforward, and affordable on systems of moderate size when truncated at single and double excitations. By adding a perturbative correction for triple excitations, coupled cluster theory can generate results within chemical accuracy for weakly correlated systems. Unfortunately, its successes can not be extended directly to the region of strong correlation.

Strong correlation is associated with quasi-degeneracy near the Fermi level and with the dominance of the two-body part of the Hamiltonian over the one-body part.  Under such conditions, the mean-field picture is qualitatively incorrect, and therefore perturbative methods based upon the mean-field reference break down.  Coupled cluster theory is not immune to this failure.  Typically, coupled cluster methods overcorrelate in strongly-correlated problems (as does single-reference perturbation theory).  One can remedy these failures by including higher-order cluster operators, but there may not be a clean truncation scheme as the magnitude of the cluster amplitudes may not decay as the excitation level increases.\cite{Degroote2016} 

Were strong correlation rare, it might perhaps be not too serious a problem.  Unfortunately, it is pervasive, and occurs, for example, in closely-spaced excited states, in bond formation and breaking (i.e. the ``recoupling region'') and at the dissociation limit, or when the system contains transition metals with partially filled $d$ or $f$ shells.  

One way to address the issue of strong correlation is to allow for the mean-field reference to spontaneously break symmetry.\cite{SymmetryDilemma}  Often, this will provide a reasonably accurate energy, at the cost of a somewhat unphysical wave function.  The unphysical nature of broken-symmetry wave functions can cause difficulties in assigning states, and can lead to significant errors in the prediction of properties related to the symmetry which has been broken.  One is then forced to choose between having more accurate total energies on the one hand, and correct symmetry properties on the other.  This is the well-known symmetry dilemma.

At the mean-field level, the symmetry dilemma is, we believe, essentially solved by the symmetry-projected Hartree-Fock (PHF) method.\cite{Lowdin55c,Ring80,Blaizot85,Schmid2004,PHF} In PHF theory, only those portions of a broken-symmetry determinant which have the correct symmetry are retained.  This is accomplished through the use of a symmetry projection operator.  By representing the PHF wave function as a linear combination of degenerate and non-orthogonal broken-symmetry determinants obtainable one from another by symmetry rotation operators, the PHF energy and properties can be obtained at a computational scaling comparable to that of the underlying broken-symmetry Hartree-Fock.\cite{PHF}  By construction, the wave function retains the correct symmetries, but it inherits the energetic advantages of broken-symmetry mean-field methods.

While PHF successfully resolves the symmetry dilemma at the mean-field level, and often offers a reasonably accurate description of strong correlations, it does not account for the remaining weak correlations.  Ultimately, the symmetry-projected mean-field state is insufficiently complete, and PHF must be combined with some other technique to account for dynamic correlations.  Unfortunately, this is far from simple.  While a great deal of effort has been put into combining symmetry projection and coupled cluster theory, the optimal approach is not entirely clear at present.

One major avenue of investigation has been to work with a perturbative or coupled-cluster style correction of the PHF wave function where the correlation operators are themselves symmetry adapted.\cite{Degroote2016,WahlenStrothman2016,Qiu2016,Qiu2017,Gomez2017,Hermes2017}  These approaches benefit from the observation\cite{Piecuch1996,Qiu2016,Qiu2017,SGCCSD} that the PHF wave function often has a rather simple particle-hole representation in the symmetry-adapted basis, even though its particle-hole expansion in the broken-symmetry basis may be complicated.  One may then interpolate between PHF and coupled cluster,\cite{Degroote2016,Gomez2017} or attempt to add a symmetry-adapted cluster operator directly atop the PHF wave function.\cite{WahlenStrothman2016,Qiu2017}

Alternatively, and more closely akin to PHF, one could instead write the wave function by acting a symmetry projection operator on a broken-symmetry coupled cluster state.\cite{Duguet2015,Qiu2017b,Tsuchimochi2017}  It is this latter approach that we take here.  We will focus on spin symmetry in this manuscript, because it is the symmetry that spontaneously breaks in molecular systems, but the basic framework can be easily generalized to other continuous symmetries, such as number symmetry.\cite{Duguet2015,Duguet2016,Qiu2017b,DuguetNew}

In previous work, we proposed a formalism we referred to as the disentangled cluster approximation to help overcome the difficulties associated with calculating overlaps of nonorthogonal coupled cluster wave functions.  This we do by working explicitly with particle-hole excitations out of the broken-symmetry reference, which requires us to introduce an auxiliary cluster operator which we refer to as the disentangled cluster operator.  Systematic approximations to this disentangled cluster operator provide a way to approach the exact symmetry projection.  We should note that very similar ideas were introduced in Ref. \onlinecite{Duguet2015}, albeit with a different means of solving for the broken-symmetry cluster operator which led to less accurate results.  A forthcoming publication\cite{DuguetNew} will explore the similarities and differences between the two techniques in detail.

In both the mean-field and the correlated cases, the symmetry projection can be done in a \textit{post hoc} manner by first optimizing a broken-symmetry state and then applying the symmetry projection operator after the fact, or in a more integrated manner by optimizing the broken-symmetry state in the presence of the projection operator.  To be consistent with the terminology used in PHF theory, we refer to the former approach as projection after variation (PAV) though it may be more appropriate to call it projection after optimization; the latter approach is known as variation after projection (VAP) and is the focus of this manuscript.  That is, while our previous work focused on the formal details of the disentangled cluster approximation and many of our results were generated within the PAV approach, here we are specifically interested in assessing how the presence of the symmetry projection operator affects the broken-symmetry coupled cluster state and how that, in turn, affects the quality of the calculation.

Let us take a moment to be clear about the reference determinant.  The PAV approach to symmetry-projected mean-field can lead to rather alarming potential energy curves with discontinuous derivatives reflecting the underlying fact that for not all Hamiltonian parameters can a broken symmetry Hartree-Fock solution be obtained.  The VAP approach remedies this deficiency.  Throughout this work, we thus obtain the reference determinant of the calculation from the VAP projected Hartree-Fock calculation.  When we refer to VAP and PAV for coupled cluster, we mean to refer explicitly to a second choice: regardless of the way in which the reference determinant is obtained, one could obtain the cluster amplitudes in the presence of the projection operator (VAP) or without it (PAV).  Our goal here is to discuss the impact of this second choice; the mean-field reference, we believe, should always be obtained via its own VAP procedure when that is possible.

\section{Theory}
Because symmetry-projected coupled cluster is fairly new, we think it wise to begin with a relatively quick review of the basic theory.  A more detailed discussion can be found in Ref. \onlinecite{Qiu2017b}.

The basic idea of projected coupled cluster is to insert the wave function ansatz
\begin{equation}
|\Psi_\mathrm{PCC}\rangle = P \, \mathrm{e}^U |\Phi\rangle
\end{equation}
into the Schr\"odinger equation.  Here, $|\Phi\rangle$ is some broken-symmetry reference determinant, $U$ is a cluster operator creating excitations from it, and $P$ is a projection operator restoring some or all of the symmetries broken in $|\Phi\rangle$.  The energy expression and amplitude equations are obtained by left-multiplying by a set of broken-symmetry determinants, and we see that
\begin{subequations}
\begin{align}
0 &= \langle \Phi| \left(H-E\right) \, P \, \mathrm{e}^U |\Phi\rangle,
\\
0 &= \langle \mu| \left(H - E\right) \, P \, \mathrm{e}^U |\Phi\rangle,
\end{align}
\label{Eqns:PUCCBase}
\end{subequations}
where $\langle \mu|$ stands for an excited determinant.  We use the first equation to define the energy, and solve the second set of equations to solve for the cluster amplitudes in $U$.

The presence of the projection operator creates significant complications in evaluating the right-hand-sides of Eqn. \ref{Eqns:PUCCBase}.  It proves helpful to write the projection operator in an integral representation,\cite{Schmid2004,PHF} generically as
\begin{equation}
P = \int \mathrm{d}\Omega \, w(\Omega) \, R(\Omega)
\end{equation}
where $w(\Omega)$ is a (normalized) weight and $R(\Omega)$ is a one-body symmetry rotation operator.  In the case of spin projection onto a singlet, for example, we have explicitly
\begin{equation}
P = \int_0^{2\pi} \mathrm{d}\alpha \, \int_0^\pi \sin(\beta) \, \mathrm{d}\beta \, \int_0^{2\pi} \mathrm{d}\gamma \, w(\Omega) \, R(\Omega)
\end{equation}
where the weight and rotation operator are respectively
\begin{subequations}
\begin{align}
w(\Omega) &= \frac{1}{8 \pi^2},
\\
R(\Omega) &=  \mathrm{e}^{-\mathrm{i} \, \alpha \, S_z} \, \mathrm{e}^{-\mathrm{i} \, \beta \, S_y} \, \mathrm{e}^{-\mathrm{i} \, \gamma \, S_z}.
\end{align}
\end{subequations}
Other symmetries (or other quantum numbers of the same symmetry) would use different weights and rotation operators.

Inserting our form for the projection operator into the working equations, we obtain
\begin{subequations}
\begin{align}
0 &= \int \mathrm{d}\Omega \, w(\Omega) \, \langle \Phi| R(\Omega) \, \left(H-E\right) \, \mathrm{e}^U |\Phi\rangle,
\\
0 &= \int \mathrm{d}\Omega \, w(\Omega) \, \langle \mu | R(\Omega) \, \left(H-E\right) \, \mathrm{e}^U |\Phi\rangle.
\end{align}
\end{subequations}
We have used the fact the projection operator commutes with the Hamiltonian.  Ultimately, then, we must evaluate norm and energy kernels
\begin{subequations}
\begin{align}
\mathcal{N}(\Omega) &= \langle\Phi| R(\Omega) \, \mathrm{e}^{U} |\Phi\rangle,
\\
\mathcal{N}(\Omega) \, \mathcal{H}(\Omega) &= \langle\Phi| R(\Omega) \, H \, \mathrm{e}^{U} |\Phi\rangle,
\\
\mathcal{N}(\Omega) \, \mathcal{N}_{\mu}(\Omega) &= \langle\mu| R(\Omega) \, \mathrm{e}^{U} |\Phi\rangle,
\\
\mathcal{N}(\Omega) \, \mathcal{H}_{\mu}(\Omega) &= \langle\mu| R(\Omega) \, H \, \mathrm{e}^{U} |\Phi\rangle.
\end{align}
\end{subequations}
With these objects in hand, we can write the energy and amplitude equations in terms of weighted integrals over the various kernels.

Unfortunately, while these kernels can be readily evaluated in the Hartree-Fock limit where the cluster operator $U$ is disregarded, their evaluation is rather cumbersome in general.  To solve this difficulty we use what we refer to as the disentangled cluster approximation.  By the Thouless theorem,\cite{Thouless1960} the rotation operator when acting on a single determinant to the left can be replaced by the exponential of a single de-excitation operator:
\begin{equation}
\langle \Phi| R(\Omega) = \langle \Phi | R(\Omega) |\Phi\rangle \, \langle \Phi| \mathrm{e}^{V_1(\Omega)}.
\end{equation}
We can make a similar replacement for the excited kernels:
\begin{subequations}
\begin{align}
\langle \mu| R(\Omega)
 &= \langle \Phi| \mathcal{Q}_\mu \, R(\Omega)
\\
 &= \langle \Phi| R(\Omega) \, R^{-1}(\Omega) \, \mathcal{Q}_\mu \, R(\Omega)
\\
 &= \langle \Phi | R(\Omega) |\Phi\rangle \, \langle \Phi| \mathrm{e}^{V_1(\Omega)} \, R^{-1}(\Omega) \, \mathcal{Q}_\mu \, R(\Omega)
\\
 &= \langle \Phi | R(\Omega) |\Phi\rangle \, \langle \Phi| \tilde{\mathcal{Q}}_\mu(\Omega) \, \mathrm{e}^{V_1(\Omega)}.
\end{align}
\end{subequations}
Here, $\mathcal{Q}_\mu$ is the deexcitation operator which converts $\langle \Phi|$ into $\langle \mu|$ and $\tilde{\mathcal{Q}}_\mu$ is its $\Omega$-dependent similarity transformation,
\begin{equation}
\tilde{\mathcal{Q}}_\mu(\Omega) = \mathrm{e}^{V_1(\Omega)} \, R^{-1}(\Omega) \, \mathcal{Q}_\mu \, R(\Omega) \, \mathrm{e}^{-V_1(\Omega)}.
\end{equation}
Suppressing the $\Omega$-dependence for brevity, the various kernels thus become 
\begin{subequations}
\begin{align}
\mathcal{N} &= \langle \Phi| R | \langle \Phi \rangle \, \langle \Phi| \mathrm{e}^{V_1} \, \mathrm{e}^{U} |\Phi\rangle,
\\
\mathcal{N} \, \mathcal{H} &= \langle \Phi| R | \Phi \rangle \, \langle \Phi| \mathrm{e}^{V_1} \, H \, \mathrm{e}^{-V_1} \, \mathrm{e}^{V_1} \, \mathrm{e}^U |\Phi\rangle,
\\
\mathcal{N} \, \mathcal{N}_{\mu} &= \langle \Phi| R |\Phi\rangle \, \langle \Phi| \tilde{\mathcal{Q}_\mu} \, \mathrm{e}^{V_1} \, \mathrm{e}^U | \Phi\rangle,
\\
\mathcal{N} \, \mathcal{H}_{\mu} &= \langle \Phi| R |\Phi\rangle \, \langle \Phi| \tilde{\mathcal{Q}_\mu} \, \mathrm{e}^{V_1} \, H \, \mathrm{e}^{-V_1} \, \mathrm{e}^{V_1} \, \mathrm{e}^U |\Phi\rangle.
\end{align}
\end{subequations}
In the Hamiltonian kernels we have introduced unity in the form of a product of exponentials, and we see that we can define a $V_1$-transformed Hamiltonian $\bar{H}_V$ as simply
\begin{equation}
\bar{H}_V = \mathrm{e}^{V_1} \, H \, \mathrm{e}^{-V_1}
\end{equation}
to simplify our expressions.

The key quantity is clearly the transformed wave function
\begin{equation}
|\Psi\rangle = \mathrm{e}^{V_1} \, \mathrm{e}^U |\Phi\rangle.
\end{equation}
This wave function can be written in a full coupled cluster expansion:
\begin{equation}
|\Psi\rangle = \mathrm{e}^W |\Phi\rangle.
\end{equation}
Because the wave function is not necessarily in intermediate normalization, we must include a zero-body cluster operator in our definition of $W$.

To obtain the disentangled cluster operator $W$, we solve a differential equation.  First, we note that the CI coefficients $\mathcal{C}$ in the expansion of $|\Psi\rangle$ can be obtained by left-projection of $|\Psi\rangle$ against $\langle \Phi|$ and particle-hole excitations out of it:
\begin{subequations}
\begin{align}
\mathcal{C}_0 &=  \langle \Phi|\Psi\rangle,
\\
\mathcal{C}_\mu &= \langle \mu| \Psi\rangle.
\end{align}
\end{subequations}
Note that from $W$ we can determine $\mathcal{C}$ and vice-versa.  Differentiating both sides of the foregoing equations with respect to the angle $\Omega$ allows us to solve a set of coupled differential equations for $W$.  To see this, note that the only $\Omega$-dependence in $W$ is carried by $V_1$.  Defining
\begin{equation}
X = \frac{\mathrm{d}V_1}{\mathrm{d}\Omega}
\end{equation}
and noting that $X$ and $V_1$ commute (because $V_1$ is a pure de-excitation operator and the $\Omega$-dependence is carried strictly by the amplitudes), we obtain
\begin{equation}
\frac{\mathrm{d} \mathcal{C}_\mu}{\mathrm{d}\Omega}
 = \langle \mu| \frac{\mathrm{d}\hfill}{\mathrm{d}\Omega} \, \mathrm{e}^{V_1} \, \mathrm{e}^U |\Phi\rangle
 = \langle \mu| X \, \mathrm{e}^W |\Phi\rangle.
\end{equation}
For example, we find that
\begin{subequations}
\begin{align}
\frac{\mathrm{d}W_0}{\mathrm{d}\beta} &= (X \, W_1)_c,
\\
\frac{\mathrm{d}W_1}{\mathrm{d}\beta} &= (\frac{1}{2} \, X \, W_1^2)_c + (X \, W_2)_c
\\
\frac{\mathrm{d}W_2}{\mathrm{d}\beta} &= (X \, W_1 \, W_2)_c + (X \, W_3)_c,
\end{align}
\label{Eqn:WDeriv}
\end{subequations}
where the subscript ``c'' denotes that only connected terms are present.  We can integrate this set of coupled differential equations to obtain the cluster amplitudes in $W$, starting from
\begin{subequations}
\begin{align}
W_0(\Omega = 0) &= 0,
\\
W_n(\Omega = 0) &= U_n,
\end{align}
\end{subequations}
where $W_n$ and $U_n$ are the $n$-body cluster operators in $W$ and $U$, respectively.

Thus far, everything we have written is exact.  For a given truncation of the broken-symmetry cluster operator $U$, we can solve the set of coupled differential equations for the disentangled cluster operators $W$.  Unfortunately, as Eqn. \ref{Eqn:WDeriv} suggests, the differential equation for one excitation level (say, $W_n$) couples to disentangled cluster operators of one higher excitation level ($W_{n+1}$).  Worse, even if the original cluster operator $U$ contains only double excitations, the disentangled cluster operators $W$ contain excitations to all levels.  We thus must choose some truncation of the disentangled cluster operator in addition to the truncation of the broken-symmetry cluster operator.  These two truncations need not be the same.

In this manuscript, for brevity, we will focus on results which truncate the higher-order disentangled cluster operators rather than those which approximate them in terms of lower-order cluster operators.  We will use a notation where the truncation of the disentangled cluster operator is included in parentheses.  Thus, for example, ``CCSD(SDT)'' would indicate that the $U$ operator includes $U_1$ and $U_2$ while the $W$ operator includes $W_0$, $W_1$, $W_2$, and $W_3$.  When the $U$ operator is truncated at singles and doubles (as it will be throughout this work), and $W$ retains only $W_0$, $W_1$, and $W_2$, both VAP and PAV calculations scale as $\mathcal{O}(N^6)$ where $N$ measures the system size.  Including $W_3$ increases the cost to $\mathcal{O}(N^7)$ and adding $W_4$ increases cost to $\mathcal{O}(N^8)$.

One should be aware that the VAP calculations are significantly more expensive than are the PAV ones.  In the PAV case, we solve standard CC equations, and then finally have a single projection step.  For the VAP case, the cost of the solution of the amplitude equations is multiplied by the number of grid points used in the projection.  Additionally, the spin rotation operator takes an unrestricted determinant which is still an eigenfunction of $S_z$ into a rotated determinant which is not.  Thus, it is simplest to use a generalized Hartree-Fock-based approach, which further increases the computational cost.  The cost of a VAP projected unrestricted coupled cluster is more than an order of magnitude higher than that of the unprojected calculation, though we note that we have been conservative with numerical integration grids needed to do the projection, and one should be able to use much smaller grids without unduly compromising the accuracy since the coupled cluster wave function is generally less symmetry broken than is its mean-field reference.  Moreover, the same sorts of techniques used to reduce the cost of standard coupled cluster may be applied to projected coupled cluster as well.

\section{Algorithm}
We have described thus far how we obtain the coupled cluster energy and residuals, and we have stated that we adjust the $U$ operator to make the residuals vanish.  To be complete, we should describe how this is actually done.  We use a quasi-Newton approach, in which we solve
\begin{equation}
0 = \mathcal{R}_\mu[U] + \mathcal{J}_{\mu\nu} \, \delta U_\nu
\end{equation}
where $\mathcal{R}_\mu$ is the residual, $\mathcal{J}_{\mu\nu}$ is some approximate Jacobian, and $\delta U_\nu$ is the update to the $\nu^\textrm{th}$ component of $U$.

The full Jacobian is of course far too large to work with.  Accordingly, we use a modification of Broyden's method\cite{Broyden1,Broyden2} to solve the amplitude equations.  The coupled cluster amplitudes are updated according to
\begin{equation}
U^{(n+1)} = U^{(n)} - J_0^{-1} \, \mathcal{R}^{(n)} - \sum_{i=1}^{n-1} \gamma_{ni} \, y^{(i)}
\label{TMH1}
\end{equation}
where $J_0^{-1}$ is our initial guess for the inverse Jacobian and $\mathcal{R}^{(n)}$ is the residual at the $n^\textrm{th}$ iteration.  The vector $y^{(i)}$ is given by
\begin{equation}
y^{(i)} = -J_0^{-1} \, \Delta \mathcal{R}^{(i)} + \Delta U^{(i)}
\label{TMH2}
\end{equation}
where
\begin{subequations}
\begin{align}
\Delta \mathcal{R}^{(i)} &= \frac{\mathcal{R}^{(i+1)} - \mathcal{R}^{(i)}}{|\mathcal{R}^{(i+1)} - \mathcal{R}^{(i)}|},
\\
\Delta U^{(i)} &=  \frac{U^{(i+1)} - U^{(i)}}{|\mathcal{R}^{(i+1)} - \mathcal{R}^{(i)}|}.
\end{align}
\end{subequations}
Finally, the matrix $\gamma$ is given as
\begin{equation}
\gamma_{ni} = \sum_{k=1}^{n-1} \left(\Delta \mathcal{R}^{(n)} \cdot \mathcal{R}^{(k)}\right) \, \beta_{ki}
\end{equation}
where in turn $\beta$ is defined via
\begin{subequations}
\begin{align}
\beta &= \left(w_0^2 + a\right)^{-1},
\\
a_{ij} &= \Delta \mathcal{R}^{(i)} \cdot \Delta \mathcal{R}^{(j)}.
\end{align}
\end{subequations}
Full details can be found in the references cited.  We use $w_0 = 0.01$ and pick as the inverse Jacobian the orbital energy denominator, i.e.
\begin{subequations}
\begin{align}
\left(J_0\right)^{-1}_{ia,ia} &= \frac{1}{\epsilon_a - \epsilon_i},
\\
\left(J_0\right)^{-1}_{ijab,ijab} &= \frac{1}{\epsilon_a + \epsilon_b - \epsilon_i - \epsilon_j}
\end{align}
\label{TMH3}
\end{subequations}
where $\epsilon_p$ is a diagonal element of the semicanonical Fock matrix.

We should perhaps say a few words about our motivation for using this approach.  The usual coupled cluster iterations can be written as
\begin{equation}
\Delta \epsilon \, U_\mathrm{new} = \mathcal{R}[U_\mathrm{old}] + \Delta \epsilon \, U_\mathrm{old}
\end{equation}
and solved for $U_\mathrm{new}$.  This can be reexpressed as
\begin{equation}
0 = \mathcal{R}[U_\mathrm{old}] - \Delta \epsilon \, \delta U
\end{equation}
where $\delta U$ is the update to $U$ (i.e. $\delta U = U_\mathrm{new} - U_\mathrm{old}$), and it becomes clear that the usual coupled cluster iterations can be thought of as a quasi-Newton procedure where the Jacobian is approximated as the (diagonal) orbital energy difference $-\Delta \epsilon$.  The fact that the Jacobian is approximated undoubtedly affects the rate of convergence, but does not change the actual solution, which of course occurs at $\mathcal{R}[U] = 0$.

In the projected case, we observe that the orbital energy difference is not a sufficiently accurate approximate Jacobian, because the structure of the amplitude equations is quite complicated, the underlying variables $U$ being obscured behind the disentangled cluster operators $W$ obtained by integrating a differential equation.  We use a variant of Broyden's method simply because Broyden's method updates the approximate Jacobian iteratively in a computationally affordable manner and converges rapidly when near the solution.

\section{Gauge Modes and the Goldstone Manifold}
The set of states created by acting the symmetry rotation operator $R(\Omega)$ on a mean-field determinant is known as the Goldstone manifold.  As we have previously noted, they are degenerate and nonorthogonal.  Moreover, every state in the Goldstone manifold leads to the same projected state (up to an overall phase, at any rate).  That is, for any determinant $|\Phi\rangle$ we have
\begin{equation}
P |\Phi\rangle = P R(\Omega) |\Phi\rangle = R(\Omega) \, P | \Phi\rangle.
\label{Eqn:Invariance}
\end{equation}
This invariance has significant consequences.

Recall that the residuals we desire to compute are
\begin{equation}
\mathcal{R}_\mu = \langle \mu| P \, \left(H - E\right) \, \mathrm{e}^U |\Phi\rangle.
\end{equation}
From the invariance condition of Eqn. \ref{Eqn:Invariance} we see that equivalently
\begin{equation}
\mathcal{R}_\mu = \langle \mu| R(\Omega) \, P \, \left(H - E\right) \, \mathrm{e}^U |\Phi\rangle
\end{equation}
for any angle $\Omega$.

Generically, the rotation operator is the exponential of some symmetry operator $\mathcal{O}$:
\begin{equation}
R(\Omega) = \mathrm{e}^{\mathrm{i} \, \Omega \, \mathcal{O}}.
\end{equation}
The invariance of the residual under symmetry rotation implies that
\begin{equation}
0 = \langle \mu | \mathcal{O}^n \, P \, \left(H - E\right) \, \mathrm{e}^U |\Phi\rangle
\label{Eqn:DefInvariantMode}
\end{equation}
for any determinant $\langle \mu|$ and for any integer power $n$, provided that $\mathcal{O}$ is the symmetry operation projected by $P$.

\begin{figure*}[t]
\includegraphics[width=0.96\columnwidth]{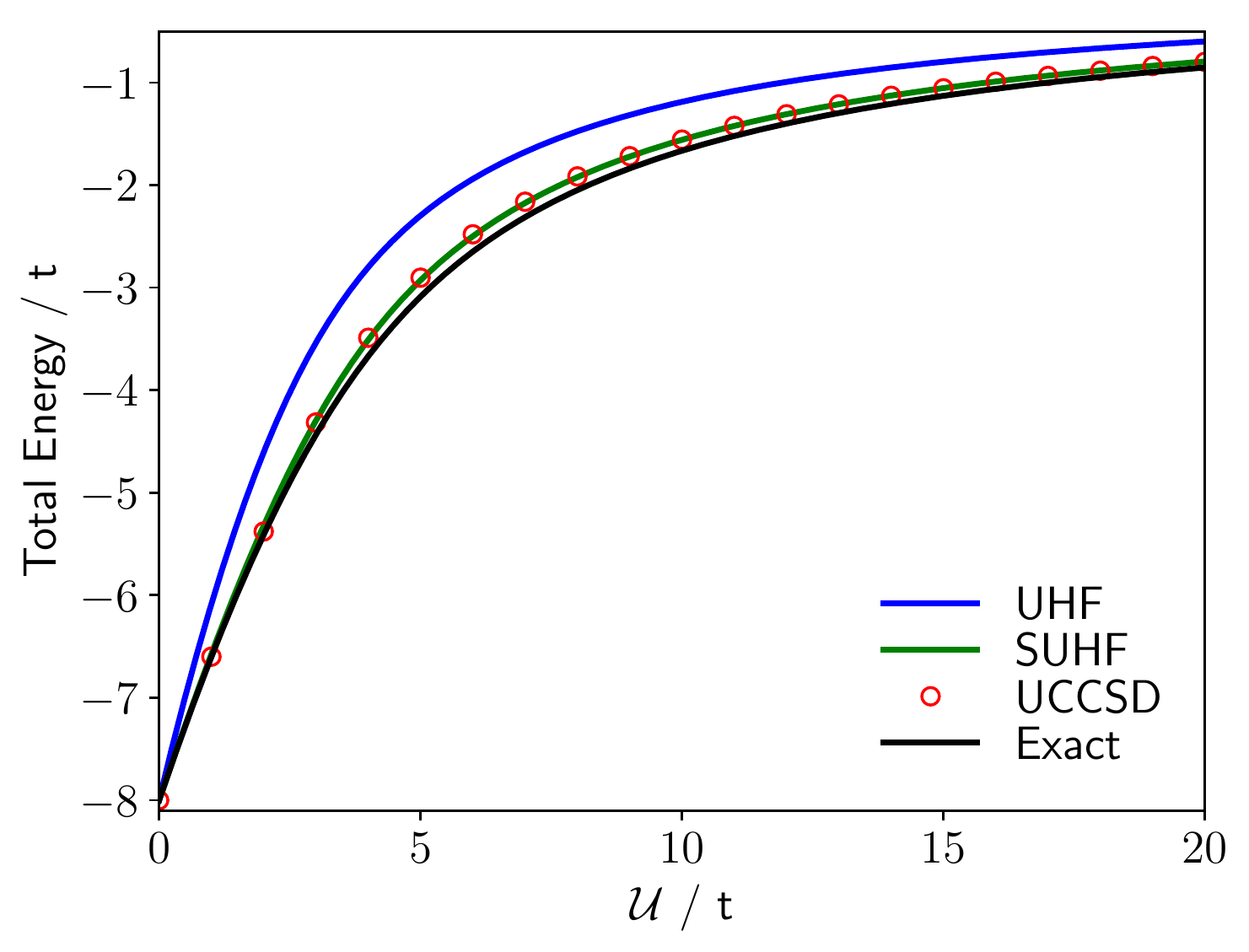}
\hfill
\includegraphics[width=0.96\columnwidth]{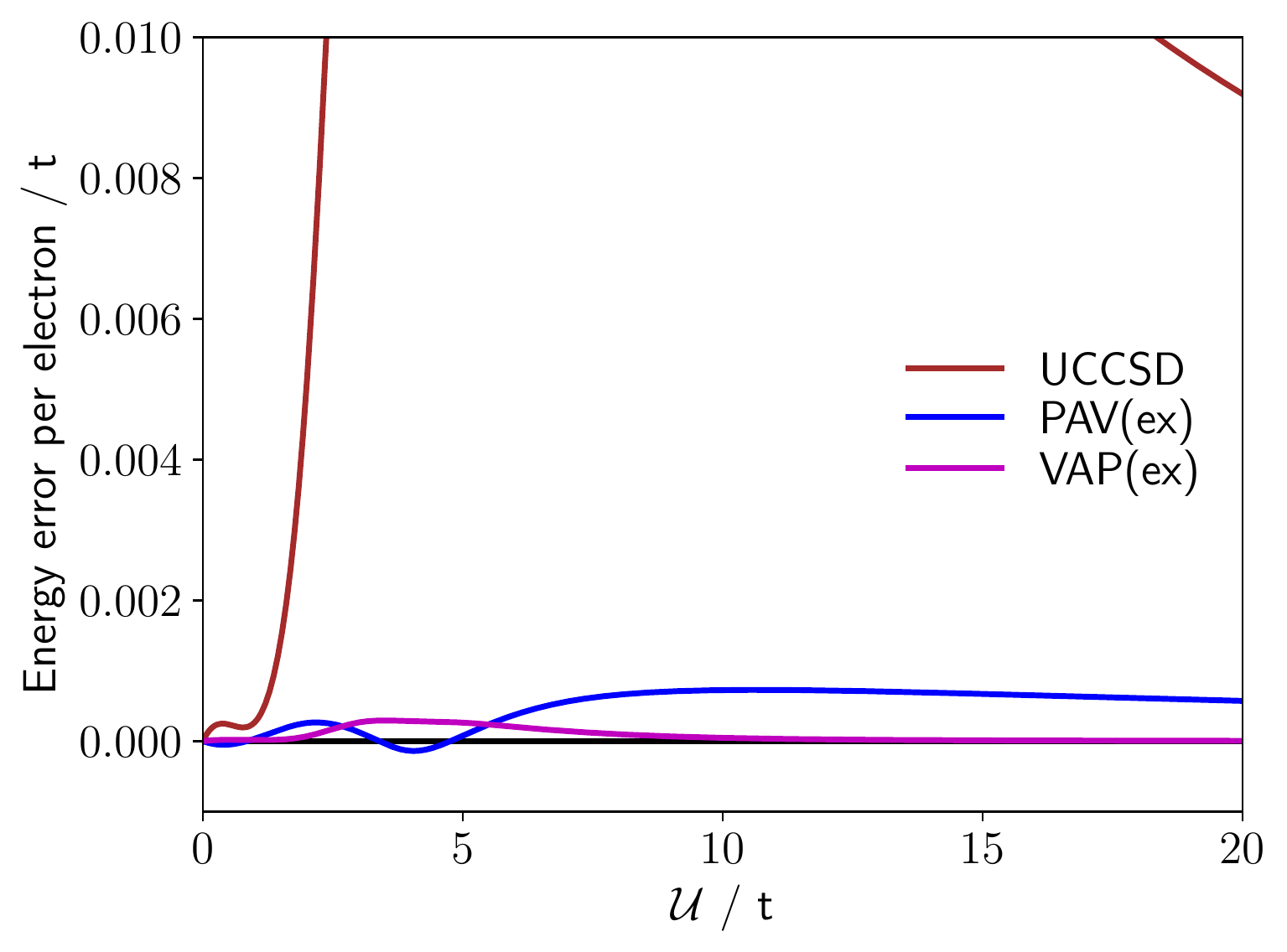}
\\
\includegraphics[width=0.96\columnwidth]{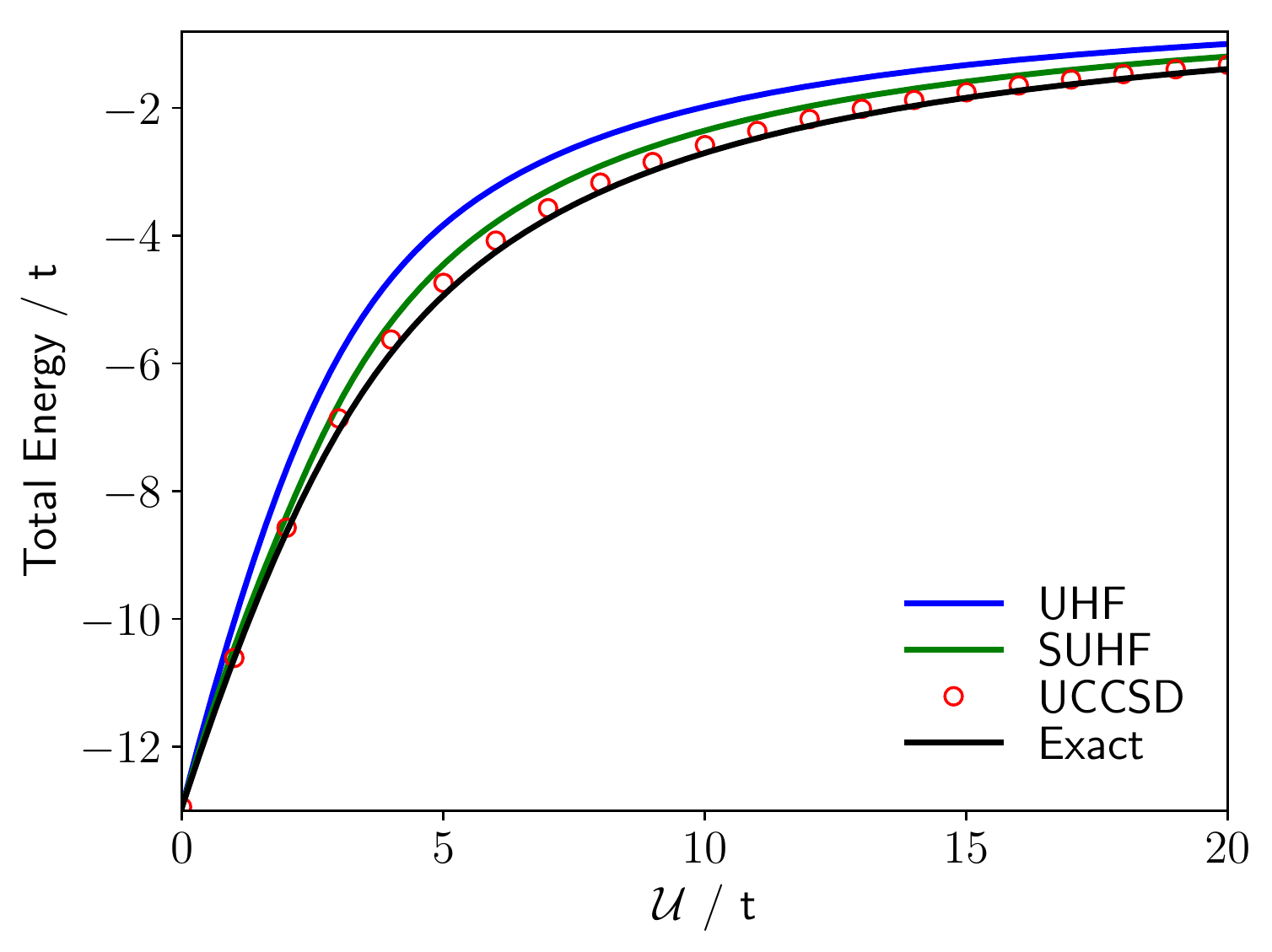}
\hfill
\includegraphics[width=0.96\columnwidth]{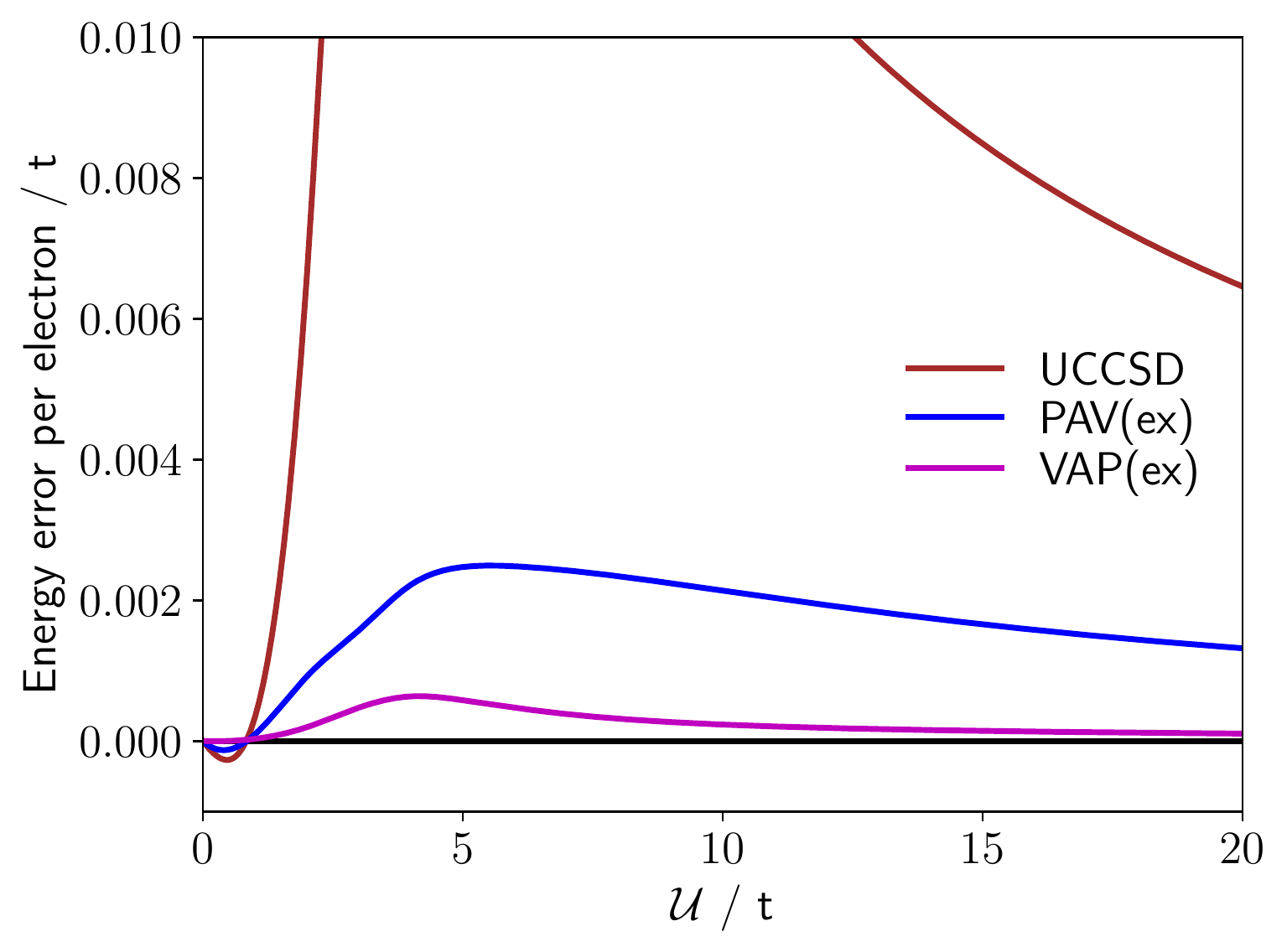}
\caption{Total energies (top row) and errors per electron (bottom row) in the periodic one-dimensional 6-site (left column) and 10-site (right columns) Hubbard lattices.
\label{Fig:Hubbard1}}
\end{figure*}

When the operator $\mathcal{O}^n$ acts on $\langle \mu|$, it creates some linear combination of determinants, so
\begin{equation}
\langle \mu| \mathcal{O}^n = L^{(n)}_{\mu,\Phi} \, \langle \Phi| + \sum L^{(n)}_{\mu\nu} \, \langle \nu|
\label{Eqn:Invariant2}
\end{equation}
where the precise details of the coefficients $L^{(n)}$ need not concern us.  What does concern is that Eqn. \ref{Eqn:DefInvariantMode} and \ref{Eqn:Invariant2} together imply that certain linear combinations of the residuals are zero simply by virtue of the projection property, and thus play no role in the determination of the cluster amplitudes $U$.  That is,
\begin{equation}
\begin{split}
0 &= L^{(n)}_{\mu,\Phi} \, \langle \Phi| P \, \left(H - E\right) \, \mathrm{e}^U |\Phi\rangle
\\
 &+ \sum_\nu L^{(n)}_{\mu,\nu} \, \langle \nu | P \, \left(H - E\right) \, \mathrm{e}^U |\Phi\rangle
\end{split}
\end{equation}
even for choices of the cluster operator $U$ which do not satisfy the amplitude equations.  We will refer to the states specified by Eqn. \ref{Eqn:Invariant2} as the gauge modes of the theory, and to remaining states as the physical modes.  The upshot of all of this discussion is that with exact projection we can separate residuals into two pieces: contributions from physical modes which should be made to vanish by adjusting $U$, and contributions from gauge modes which should vanish regardless of $U$ simply as a consequence of projection.  Only the former contain real physical information.  To put things differently, the broken-symmetry cluster operator generally has two tasks.  On the one hand it must account for dynamical correlation, while on the other it also must restore symmetry.  In the presence of the projection operator, however, the latter task becomes unnecessary, and the components of $U$ which are required to carry it out become irrelevant.

\begin{figure*}[t]
\includegraphics[width=0.96\columnwidth]{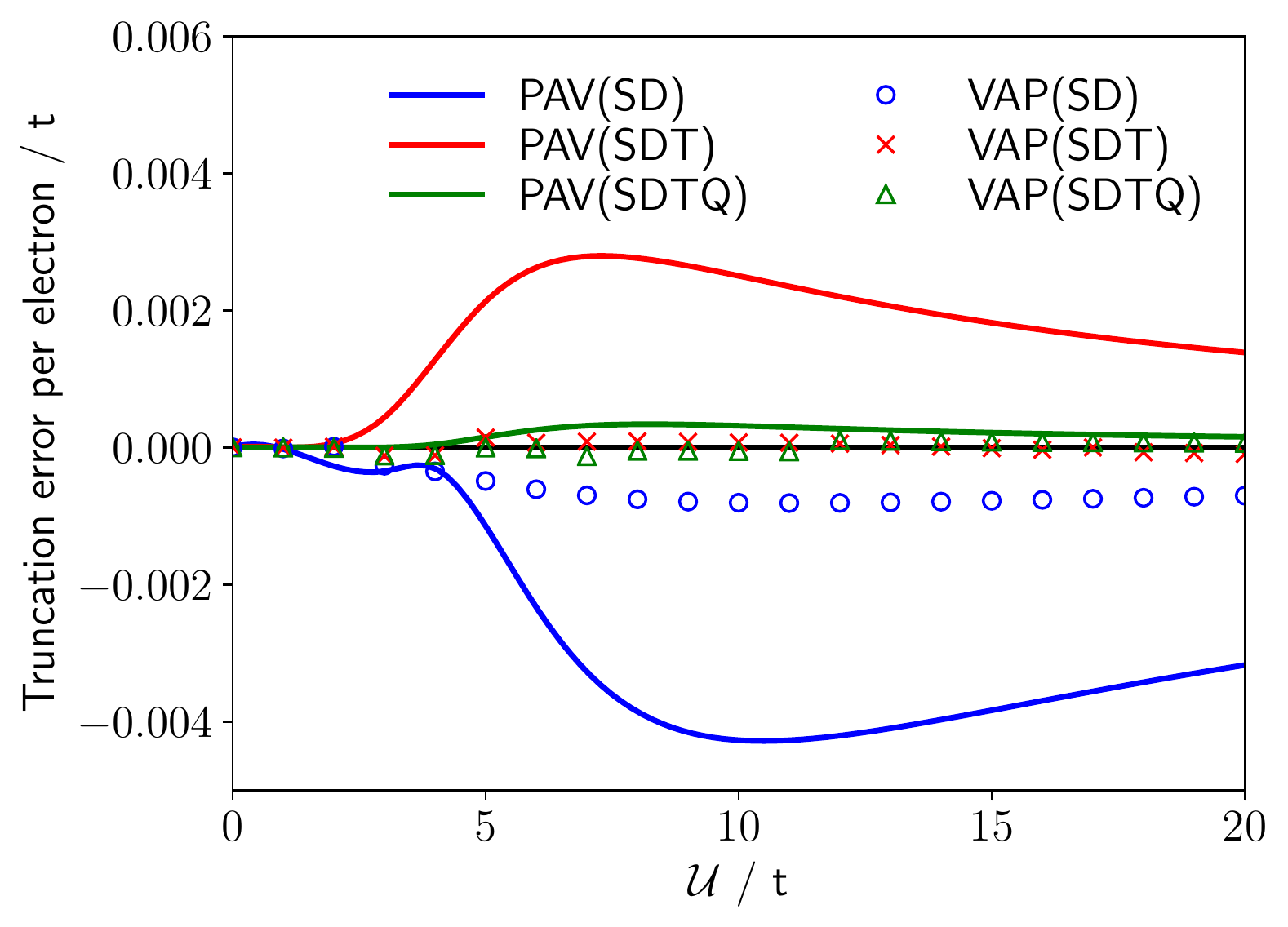}
\hfill
\includegraphics[width=0.96\columnwidth]{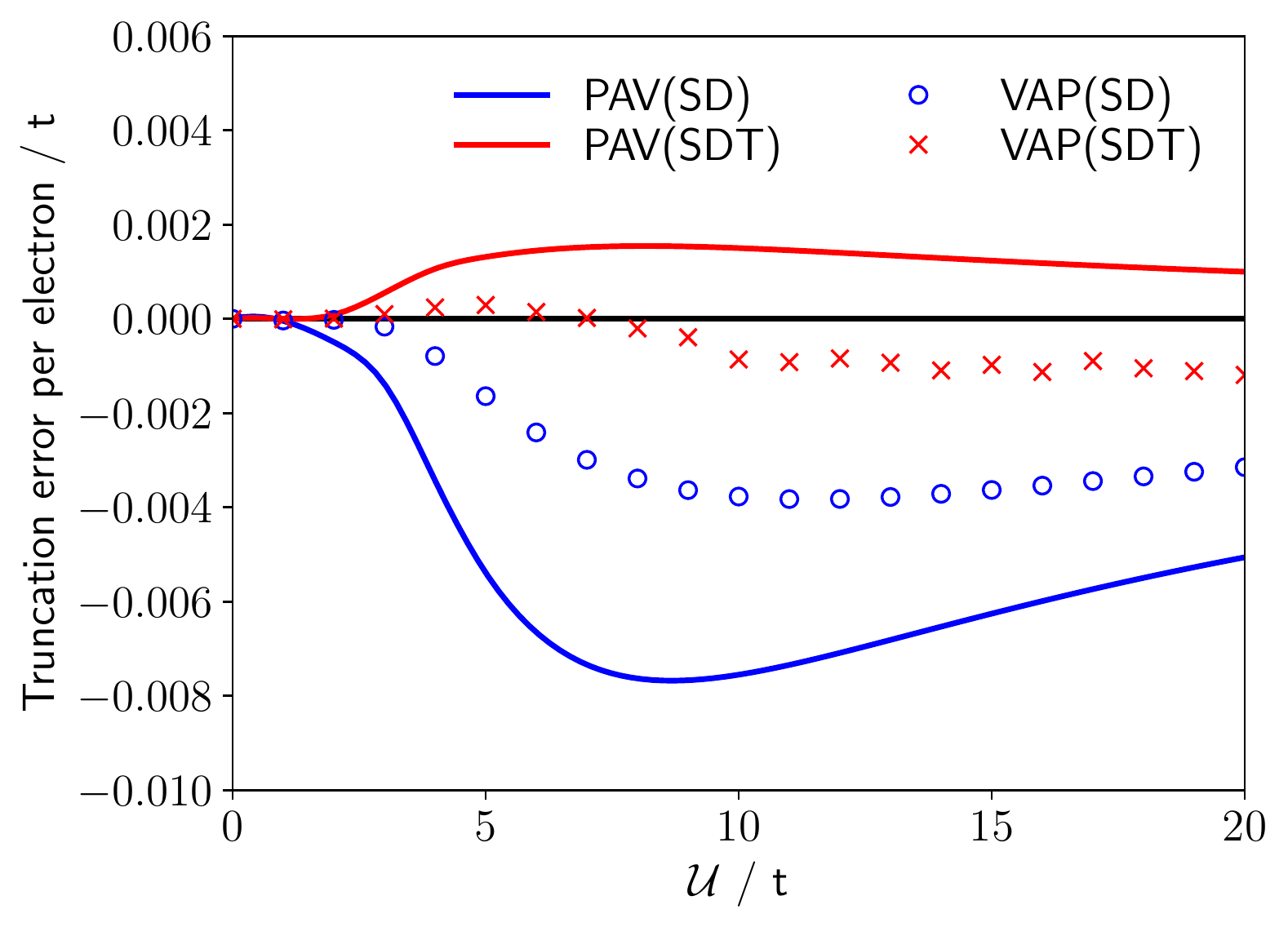}
\caption{Truncation errors per electron in the disentangled cluster approximation for the periodic one-dimension 6-site (left panel) and 10-site (right panel) Hubbard lattice.
\label{Fig:Hubbard2}}
\end{figure*}

To address convergence difficulties arising from these gauge modes, we have adopted a simple scheme.  We adjust $U$ such that the physical components of the residual go to zero.  Rather than trying to zero out the gauge components as well, we simply minimize their norm in a least-squares procedure.  Rather surprisingly, we find that minimizing the sum of squares of all residuals leads to very similar results.  Efforts to find a more satisfactory and comprehensive solution to this difficulty are underway.

\section{Results}

\begin{figure*}[t]
\includegraphics[width=0.96\columnwidth]{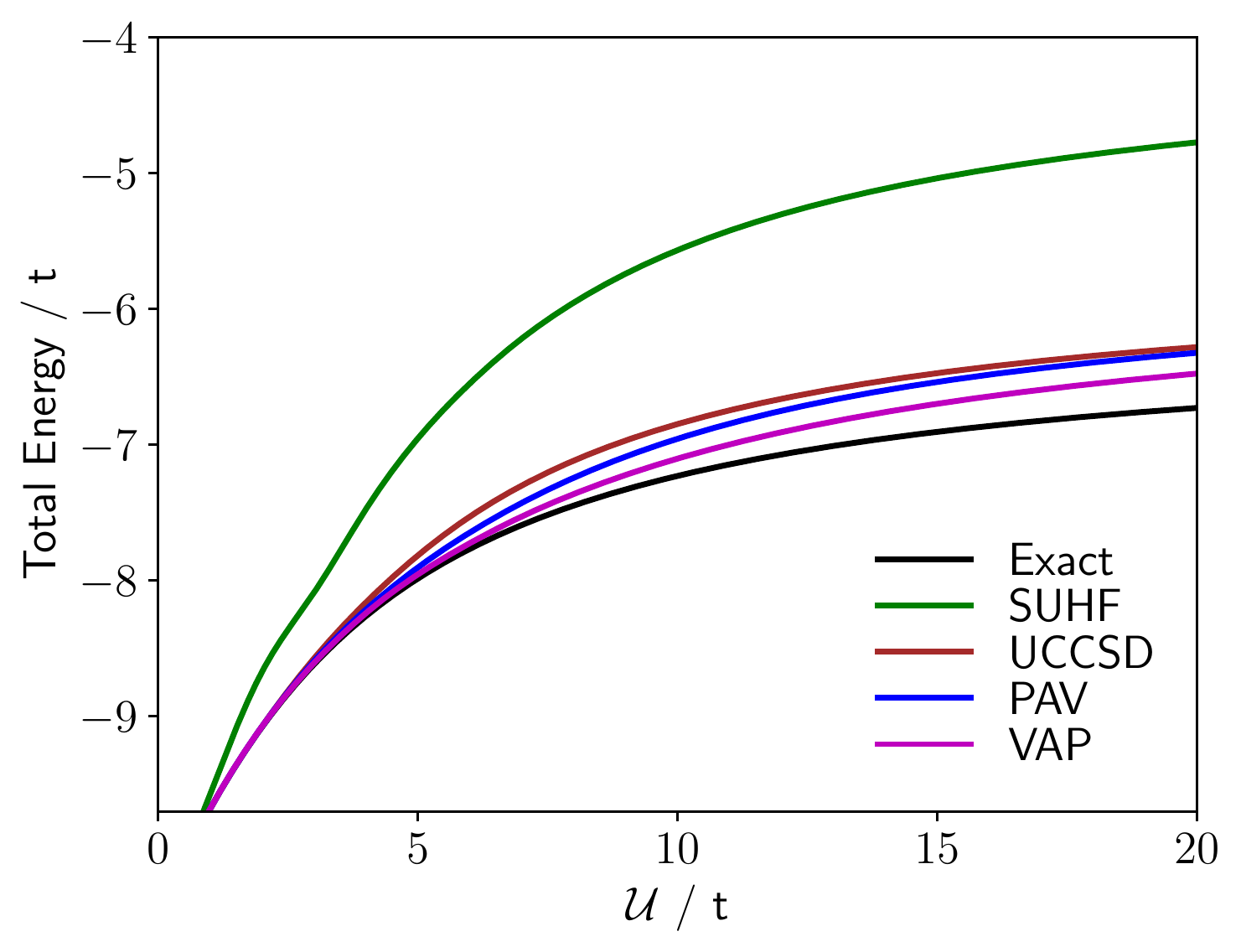}
\hfill
\includegraphics[width=0.96\columnwidth]{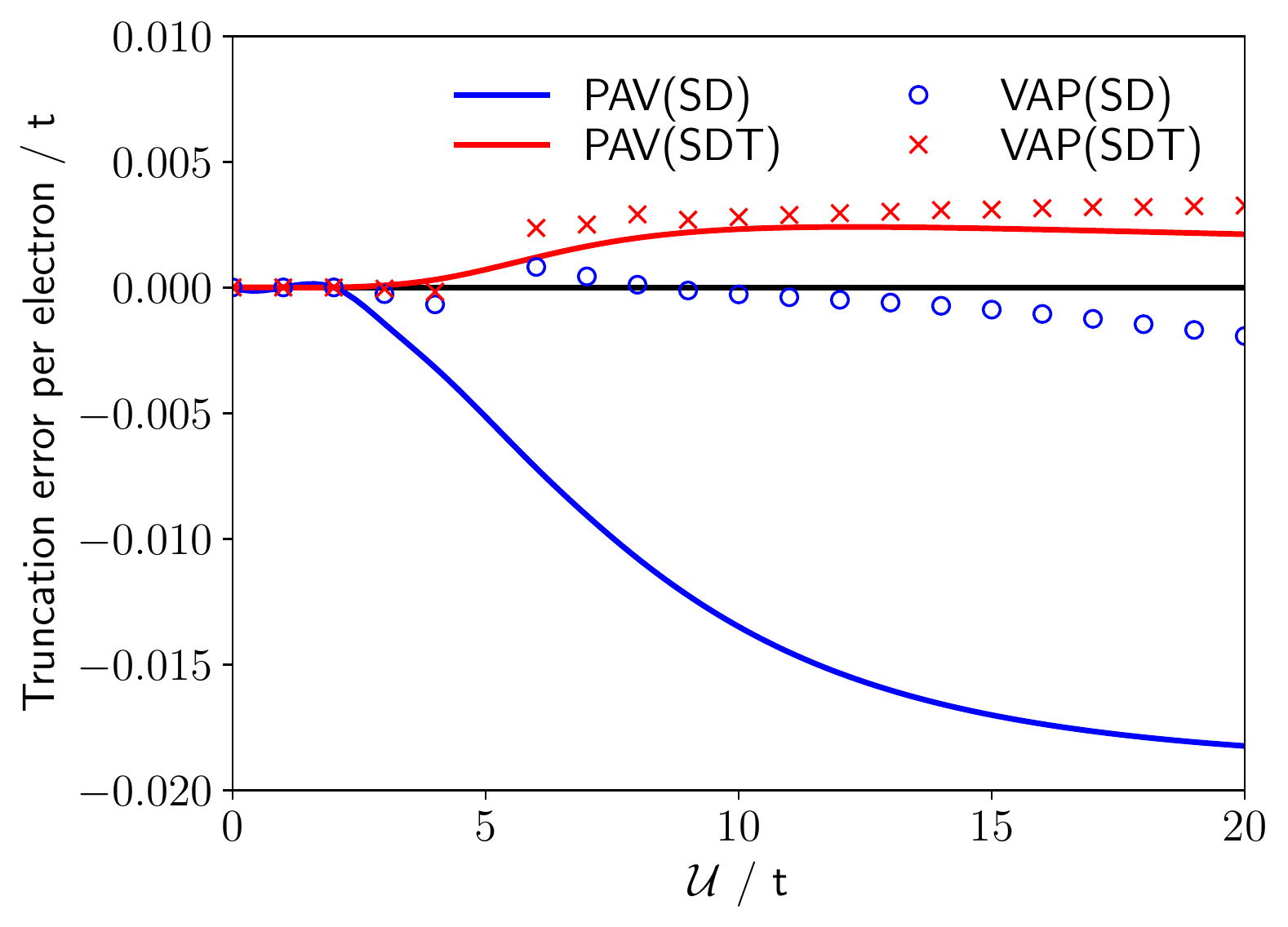}
\caption{Total energies (left panel) and truncation errors per electron (right panel) for periodic 10-site Hubbard lattices with 6 electrons.
\label{Fig:Hubbard3}}
\end{figure*}

Let us begin our assessment of the performance of symmetry-projected coupled cluster with the Hubbard Hamiltonian.\cite{Hubbard1963}  The Hamiltonian is
\begin{equation}
H = -t \sum_{\langle \mu\nu \rangle} \left(c_{\mu_\uparrow}^\dagger \, c_{\nu_\uparrow} + c_{\mu_\downarrow}^\dagger \, c_{\nu_\downarrow}\right) + \mathcal{U} \sum_\mu c_{\mu_\uparrow}^\dagger \, c_{\mu_\downarrow}^\dagger \, c_{\mu_\downarrow} \, c_{\mu_\uparrow}
\end{equation}
where $\mu$ and $\nu$ index lattice sites; the notation $\langle \mu\nu\rangle$ restricts the summation to sites connected in the lattice.  Note that the Hubbard parameter $\mathcal{U}$ is not the cluster operator $U$.  As $\mathcal{U}/t$ increases, the Hubbard model becomes more and more correlated.  We restrict our attention to periodic lattices.  While the two-dimensional lattice is of more physical interest, we will focus on one-dimensional lattices in this work.  This is for two reasons.  First, although some high-quality benchmark data is available in two dimensions,\cite{Simons} the Lieb-Wu algorithm\cite{LiebWu} provides exact one-dimensional results.  Second, one-dimensional systems are more readily accessible via a full configuration interaction (FCI) code, and we have modified just such a code to do projected coupled cluster without needing to use the disentangled cluster approximation.  This allows us to examine the convergence of results with truncated $W$ toward those with untruncated $W$.

Although our procedure is quite general, we will focus on spin projection of an unrestricted Hartree-Fock (UHF) or coupled cluster with singles and doubles (UCCSD) wave function.  We denote spin-projected versions of UHF and UCCSD as SUHF and SUCCSD, respectively.  As discussed earlier, all calculations will use the broken-symmetry determinant of VAP-SUHF as a reference, and not the variationally optimized UHF state.

Figure \ref{Fig:Hubbard1} shows total energies (top row) and errors per electron with respect to the exact result (bottom row) for the 6-site (left column) and 10-site (right column) periodic one-dimensional Hubbard lattices.  Note first that SUHF provides excellent agreement with the exact result for large enough $\mathcal{U}/t$, though convergence toward the exact result is somewhat slow.  Significantly better results are achieved with UCCSD, though it performs better for large or small $\mathcal{U}/t$ than it does in the intermediate $\mathcal{U}/t$ regime.  Simply projecting the UCCSD to give a PAV version of SUCCSD improves results significantly; solving the coupled cluster equations in the presence of the projection operator gives VAP results which are within 0.001 $t$ per electron almost everywhere.  The VAP procedure is clearly more accurate in projected coupled cluster, just as it is in projected Hartree-Fock.

On the other hand, these results have been obtained using our modified FCI code.  The disentangled cluster approximation is essential for practical calculations.  Figure \ref{Fig:Hubbard2} thus shows the convergence of the disentangled cluster approximation in SUCCSD toward the corresponding untruncated result in the same lattices as were considered in Fig. \ref{Fig:Hubbard1}.  We see that not only is VAP more accurate than PAV, but it also converges much more rapidly with truncation of the disentangled cluster operator.  The truncation error reduces to below 0.001 $t$ per electron only once we include disentangled clusters all the way through $W_4$ in PAV calculations, while with VAP we must include only up to $W_2$. 

Thus far we have only considered half-filled lattices, which are perhaps a best-case scenario for our techniques because SUHF and UCCSD are both essentially correct already for strong interactions (large $\mathcal{U}/t$).  For doped lattices, this need not be the case. Figure \ref{Fig:Hubbard3} shows results for a doped Hubbard lattice which contains 6 electrons in 10 sites.  So far as we can determine, the actual Hartree-Fock ground state here is of generalized Hartree-Fock (GHF) character and breaks both $S^2$ and $S_z$ symmetries, so presumably the optimal approach is a spin-projected GHF in combination with coupled cluster.  In this doped lattice, the SUHF still has very large errors, as does the UCCSD.  The simple PAV correction to UCCSD is not a tremendous improvement overall, while VAP decreases the errors by about a factor of two for large $\mathcal{U}/t$.  As the right panel makes clear, the effects of truncating the disentangled cluster operator are larger here than in half-filling, presumably because the cluster operator has more to do here than in the half-filled case.

\begin{figure}[t]
\includegraphics[width=\columnwidth]{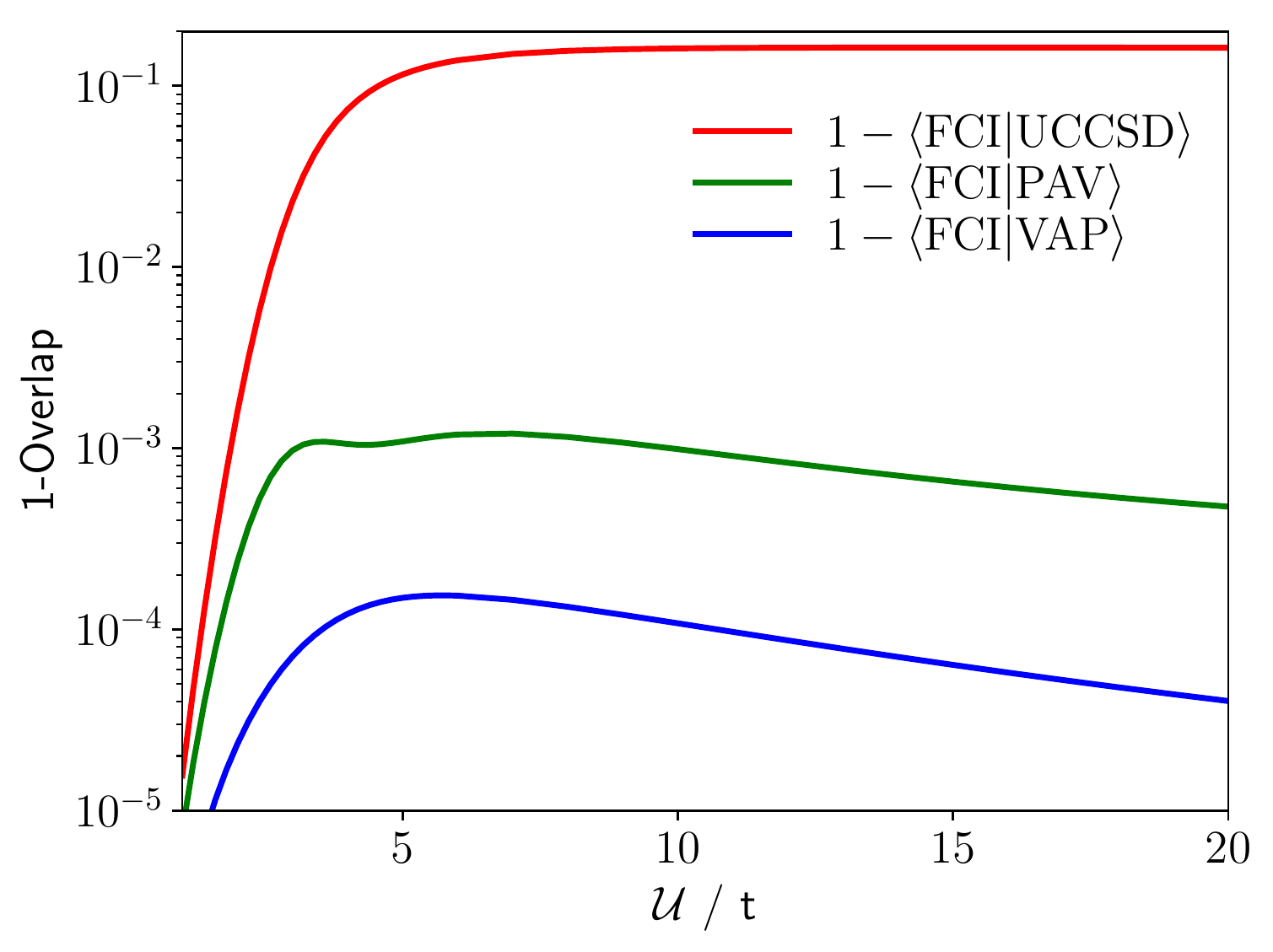}
\caption{Overlap of the UCCSD and symmetry projected UCCSD wave functions with the exact ground state in the half-filled periodic 6-site Hubbard lattice.
\label{Fig:Overlap}}
\end{figure}

Our results have focused on total energies, and the implications of symmetry projection for the properties predicted by projected coupled cluster are not yet clear.  However, we can at least check the quality of the projected coupled cluster wave function by simply taking its normalized overlap with the exact ground state.  This is shown in Fig. \ref{Fig:Overlap} for the 6-site Hubbard model.  Even though the UCCSD wave function delivers excellent energetics, it has relatively poor overlap with the exact ground state simply by virtue of its spin contamination.  The overlap of the projected UCCSD wave function with the exact ground state is nearly one for all values of $\mathcal{U}/t$. suggesting that the properties predicted by SUCCSD will be of excellent quality (at least in this system).  Overlaps with VAP are even better than those with PAV.

\begin{figure*}[t]
\includegraphics[width=0.96\columnwidth]{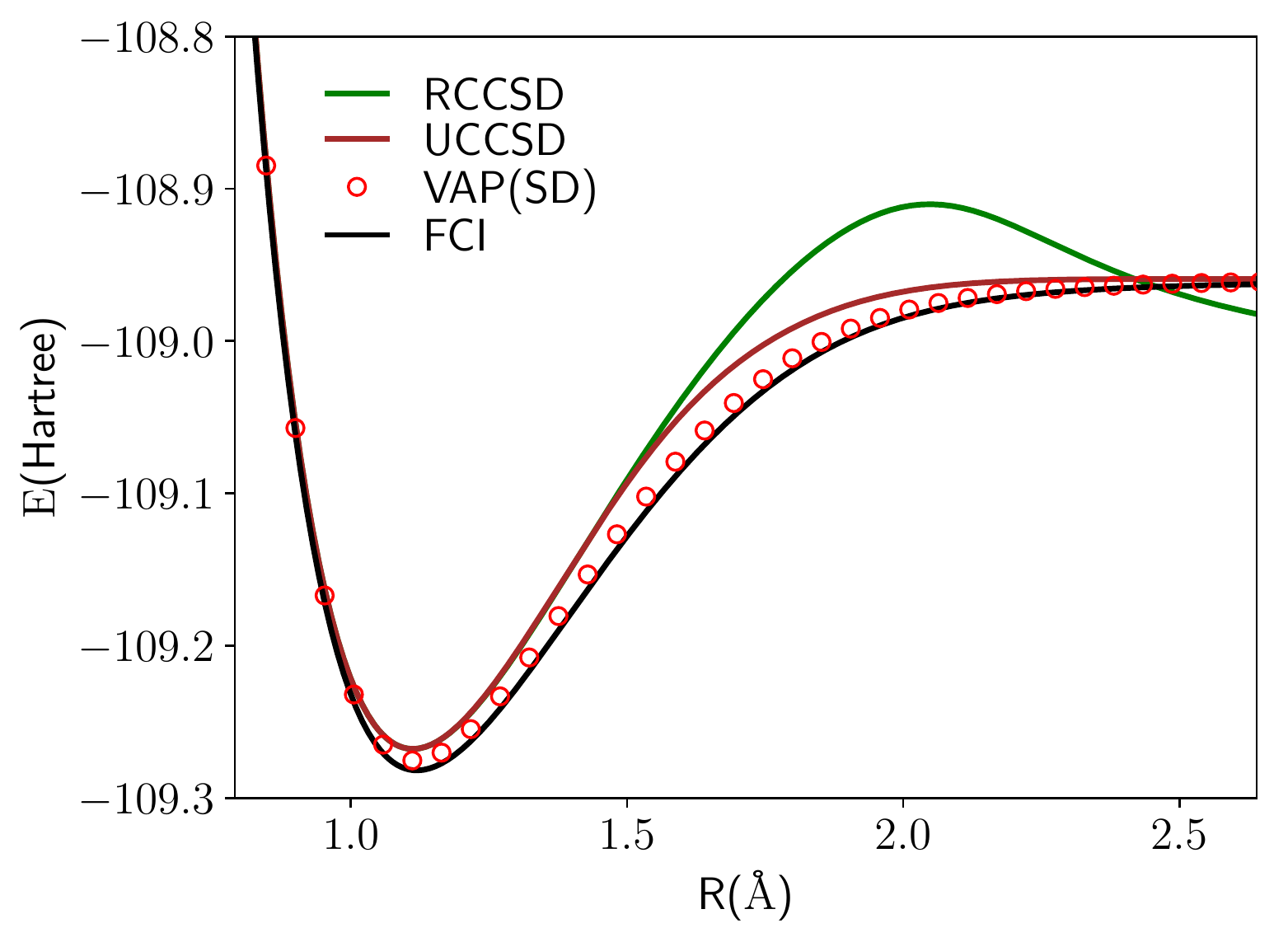}
\hfill
\includegraphics[width=0.96\columnwidth]{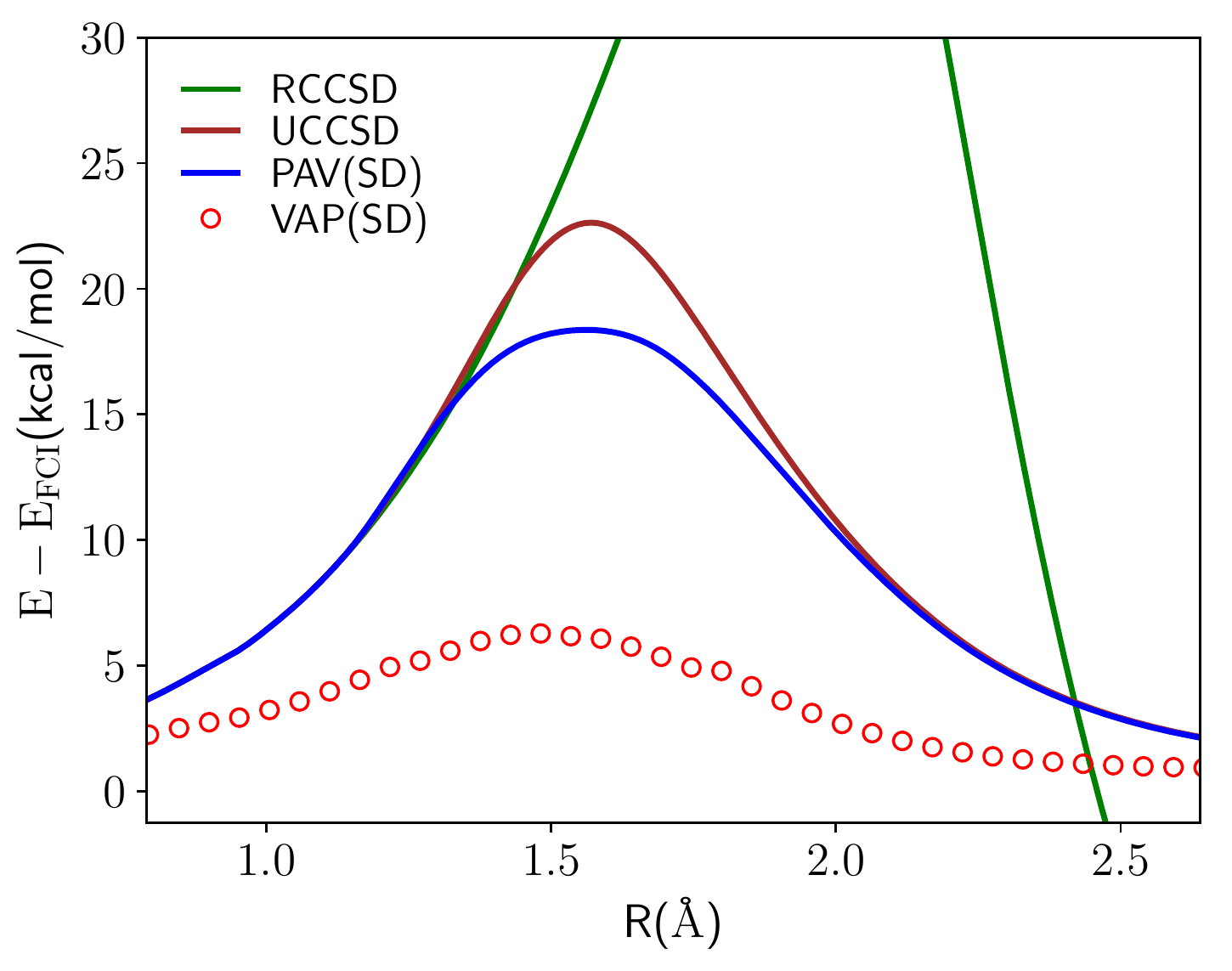}
\caption{Energies of the N$_2$ molecule using the cc-pVDZ basis set. Left panel: Total energies. Right panel: Errors with respect to full configuration interaction (FCI).
\label{Fig:N2}}
\end{figure*}

Finally, let us briefly consider a molecular system.  Figure \ref{Fig:N2} shows results for breaking the triple bond in N$_2$ in the cc-pVDZ basis set.  Although this is insufficient for quantitative accuracy, it displays the qualitative features of our technique quite well.  Clearly, restricted CCSD (RCCSD) badly overcorrelates near dissociation and has an unphysical bump.  Breaking symmetry to produce UCCSD substantially cures the problem, but with substantial remaining errors with respect to the exact FCI results.  In this case, PAV offers only modest improvement upon UCCSD itself, while VAP significantly decreases the error.  Optimizing the coupled cluster amplitudes in the presence of the projection allows the coupled cluster to focus less on restoring symmetry and more on describing dynamic correlation.

\section{Conclusion}
The fundamental difficulty with describing strong correlation is that different systems exhibit strong correlations in different ways.  This makes is difficult to come up with clever universal procedures to treat strong correlation, and typically we resort to brute force instead, by using some form of active space method in which the reference state might be expanded as a linear combination of a vast number of determinants.  The promise of symmetry projected methods is not that they deliver chemical accuracy for strongly-correlated problems, it is that they are often accurate enough to describe the basic physics in a relatively black-box manner and at a tremendously favorable mean-field computational scaling.  Neither active space methods nor symmetry projected methods are completely satisfactory, however, because neither includes a detailed description of the remaining weak correlations which are needed to deliver chemical accuracy.  In the case of active space methods, for example, one ultimately uses some form of multireference correlated technique.  The analog in the case of symmetry projection is a symmetry-projected correlated state.  These approaches are very new, but in the past few years we have begun to see progress in this area.

It would be exceedingly tempting to simply solve, for example, the projected Hartree-Fock equations to obtain a symmetry-broken determinant which could serve as the reference state for a traditional broken-symmetry coupled cluster wave function which could subsequently be projected.  This projection after variation approach, however, is not really good enough in the mean-field case and it does not appear to be good enough in the correlated case either.  Instead, it appears to be important to allow the cluster operator to relax in the presence of the projection.  This seems intuitive enough.  After all, part of what the cluster operator is asked to do in a standard broken-symmetry coupled cluster calculation is to restore symmetry (that is, the full coupled cluster on a broken-symmetry reference is a symmetry-adapted state).  By adjusting the cluster operator in the presence of the projection, we relieve it of this task and ask it only to describe the dynamic correlations to which it is most well suited.  Our results here suggest to us that projected coupled cluster in this variation after projection approach is a promising candidate for the treatment of all sorts of problems where both weak and strong correlations have an important role to play.

\begin{acknowledgments}
This work was supported by the Department of Energy under award DE-FG02-04ER15523.   G.E.S. is a Welch Foundation Chair (C-0036).  We would like to acknowledge computational support provided by the Center for the Computational Design of Functional Layered Materials, an Energy Frontier Research Center funded by the U.S. Department of Energy, Office of Science, Basic Energy Sciences under Award $\textrm{DE-SC0012575}$.
\end{acknowledgments}

\bibliography{PUCC}
\end{document}